\documentclass{article}

\usepackage{arxiv}

\usepackage[utf8]{inputenc} 
\usepackage[T1]{fontenc}    
\usepackage{hyperref}       
\usepackage{url}            
\usepackage{booktabs}       
\usepackage{amsfonts}       
\usepackage{nicefrac}       
\usepackage{microtype}      
\usepackage{graphicx}
\graphicspath{ {./figures/} }
\usepackage{subcaption}

\usepackage{amssymb}
\usepackage{mathrsfs}
\usepackage{mathtools}
\usepackage{centernot}
\usepackage{comment}
\usepackage[normalem]{ulem}
\usepackage{soul}
\usepackage{afterpage}

\hypersetup{pdfauthor={Vallarano Nicol`'o.; 
Squartini, Tiziano.; Tessone, Claudio J..}, pdftitle={Exploring the Bitcoin Mesoscale}}

\title{Exploring the Mesoscopic Structure\\of Bitcoin During its First Decade of Life}

\author{
Nicol\`o Vallarano \\
Blockchain \& Distributed Ledger Technologies Group, Department of Informatics, UZH Blockchain Center \\
University of Zurich \\
Zurich, Switzerland \\
\texttt{nicolo.vallarano@uzh.ch} \\
\And
Tiziano Squartini \\
IMT School for Advanced Studies Lucca, INDAM - National Institute for Advanced Mathematics (GNAMPA) \\
IMT School for Advanced Studies Lucca \\
Lucca, Italy \\
\texttt{tiziano.squartini@imtlucca.it} \\
\And
Claudio J. Tessone \\
Blockchain \& Distributed Ledger Technologies Group, Department of Informatics, UZH Blockchain Center \\
University of Zurich \\
Zurich, Switzerland \\
\texttt{tessone@ifi.uzh.ch} \\
}

\begin{document}

\maketitle

\begin{abstract}
The public availability of the entire history of Bitcoin transactions opens up the unprecedented possibility of studying this system at the desired level of detail. Our contribution is intended to analyse the mesoscopic properties of the Bitcoin User Network (BUN) during the first half of its history, \textit{i.e.}, across the years 2011-2018. What emerges from our analysis is that the BUN is a core-periphery structure with a certain degree of ``bow-tieness'', \textit{i.e.}, admitting the presence of a Strongly-Connected Component (SCC), an IN-component (together with some tendrils attached to it) and an OUT-component. Interestingly, the evolution of the BUN structural organisation experiences fluctuations that seem to be correlated with the presence of ``bubbles'', \textit{i.e.}, periods of price surge and decline observed throughout its entire history. Our results, thus, further confirm the interplay between structural quantities and price movements reported by previous analyses.\footnote{Originally published on Ledger Journal \url{https://ledger.pitt.edu/ojs/ledger/article/view/335}}

\end{abstract}

\section{Introduction}

Introduced in 2008 by Satoshi Nakamoto,\cite{nakamoto2008bitcoin} Bitcoin is the most widely adopted cryptocurrency. Loosely speaking, it consists of a decentralised, peer-to-peer network to which users connect for exchanging native tokens, \textit{i.e.}, the bitcoins. After having been validated by the miners---according to the consensus rules that are part of the Bitcoin protocol---each transaction becomes part of a replicated database, \textit{i.e.}, the Bitcoin blockchain.\cite{halaburda2016beyond,glaser2017pervasive} The cryptographic protocols Bitcoin rests upon aim at preventing the possibility for the same digital token to be spent more than once, in absence of a central, third party that guarantees the validity of the transactions. Remarkably, the transaction-verification mechanism allows the entire Bitcoin transaction history to be openly accessible---a feature that, in turn, allows it to be analysable in the preferred representation.\cite{nakamoto2008bitcoin,antonopoulos2017mastering} Yet, data accessibility comes at a cost: while the blockchain keeps track of each valid transaction ever issued in Bitcoin tokens, it does not record any information about the identity of the issuers. Since the inception of Bitcoin, many efforts have been thus devoted to de-anonymising transactions; in particular, heuristics  have been developed for the identification of common ownership that take advantage of the way the Bitcoin protocol handles token exchanges.\cite{harrigan2016unreasonable,ron2013quantitative,fischer2021complex,moser2022resurrecting}

Like many other systems shaped by human activities, economic and financial systems have been analysed through the lenses of network science.\cite{schweitzer2009economic} Prominent results are represented by the estimation of systemic risk in interbank networks,\cite{bardoscia2015debtrank,de2006fitness} and the topological characterisation of financial crises via the detection of early-warning signals, \textit{i.e.}, statistically significant patterns emerging during their build-up phase.\cite{Squartini:2013} The topological properties of the networked configurations induced by Bitcoin transactions have started to be investigated only recently: in Javarone and Wright (2018), the authors check the small-worldness of the so-called Bitcoin User Network;\cite{javarone2018from} in Kondor \textit{et al.} (2014) and Lin \textit{et al.} (2020), the authors highlight the tendencies towards wealth accumulation and structural centralisation, respectively characterising the so-called Bitcoin Transaction Network and Bitcoin Lightning Network;\cite{Kondor:2014,lin2020lightning} in Bovet \textit{et al.} (2019), the evolution of the local properties of four different representations of the network induced by Bitcoin transactions (\textit{i.e.}, the Bitcoin User Network and the Bitcoin Address Network, at both the daily and the weekly time scale) is explored and their relationships with price movements are investigated.\cite{bovet2019evolving}

The aim of the present work is that of inspecting the interplay between the dynamics of the Bitcoin price and that of the structural properties characterising the representation induced by the exchanges between Bitcoin users (hereafter ``BUN''): specifically, we study the emergence of \emph{i)} \emph{hubs} and \emph{motifs},\cite{barabasi2009scale,bovet2019evolving} whose statistical significance is assessed by employing properly defined benchmarks,\cite{squartini2015stationarity} and \emph{ii)} the mesoscopic structures known as ``core-periphery'' and ``bow-tie'' structures.\cite{rombach2014core,de2019detecting,khakzad2012dynamic} Interestingly, we observe co-movements of price and purely structural quantities such as reciprocity and centrality. If, on the one hand, such evidence hints at the role played by hubs in determining the level of systemic risk---by which we mean fragility towards price fluctuations---characterising Bitcoin throughout its entire history, it on the other hand also reflects the way Bitcoin's ``reputation'' has evolved throughout the years, \textit{i.e.}, from a medium of exchange to a speculative asset.

The rest of the paper is structured as follows: in Section \ref{sec:data} we describe the data and define our network representation; in Section \ref{sec:results} we define the quantities of interest for our analysis and present the main results; in Section \ref{sec:discussion} we discuss their implications.

\section{Data}\label{sec:data}

As highlighted in the Introduction, Bitcoin relies upon its blockchain: a decentralised, public ledger that records the presence and amount of transactions between users. At an abstract level, a transaction is nothing but a set of input and output addresses: the output addresses that are said to be ``unspent'' (\textit{i.e.}, not yet recorded on the ledger as input addresses) can be claimed and, therefore, spent only by the owner of the corresponding cryptographic key. This is the reason why one often speaks of Bitcoin pseudonimity (rather than anonymity): someone observing the blockchain from outside is capable of seeing all spent and unspent addresses without being able to link them to their actual owner(s). A first technique to de-pseudonymise Bitcoin prescribes to use off-chain information to link addresses to real-world entities (\textit{e.g.,} one can link an address to a merchant by looking at the merchant's Bitcoin payment channels); a second one (named ``heuristic clustering'') prescribes to rest upon the heuristics that were developed to spot out common ownership of addresses. To follow on from this, we took advantage of the existing literature on heuristic clustering to identify users---intended here as entities controlling multiple addresses (such as merchants, firms, exchanges). While this may be seen as a limitation of our analysis, it perfectly fits with our interest in modelling the evolution of Bitcoin as a whole rather than that of the single entities constituting it.

\subsection{Bitcoin Address Network (BAN)} The BAN is the simplest representation that can be constructed from the blockchain records: it is a binary, directed graph whose nodes represent addresses. The direction of links is provided by the input-output relationships defining the transactions recorded on the blockchain. More formally, the BAN is a directed network $G=(N,E)$ where any two nodes $i,j \in N$ are connected via a directed edge from $i$ to $j$---meaning $(i,j)\in E$---if they participate to the same transaction with $i$ being the input and $j$ being the output. The only degree of freedom is represented by the temporal window that can be chosen for aggregating the transactions: we have opted for the weekly time scale, \textit{i.e.}, all transactions are parsed, inputs and outputs are extracted and considered as nodes on a weekly basis.

\subsection{Bitcoin User Network (BUN)} Since the same entity can control several addresses, one can define a network of users whose nodes are clusters of addresses. Let us provide a quick overview of the literature about heuristics stemming from the work of scholars like Harrigan and Fretter (2016), Ron and Shamir (2013), Androulaki \textit{et al.} 2013, and Tasca \textit{et al.} (2018), as well as a brief description of those heuristics that have been employed here. \cite{androulaki2013evaluating,tasca2018evolution,harrigan2016unreasonable,ron2013quantitative} The first one is named \emph{multi-input heuristics} and is based upon the assumption that two (or more) addresses that are part of the same input are controlled by the same user---the reason being that the private keys of all addresses must be accessible to the creator of a transaction, in order to generate it. The second one is named \emph{change-address heuristics} and is based upon the observation that transaction outputs must be fully spent: the creator of a transaction, thus, usually also controls the output address collecting the change. More specifically, we assume that if an output address is new and the amount of bitcoins transferred to it is smaller than each input, then it must represent the change, hence belonging to the creator of the transaction. Whenever the BUN is mentioned, we refer to the representation obtained by clustering the nodes of the BAN according to a combination of the two heuristics above: more formally, the BUN is a directed graph $G=(N,E)$ where any two nodes $i,j \in N$ are connected via a directed edge from $i$ to $j$---meaning $(i,j)\in E$---if at least one transaction from one of the addresses defining $i$ to one of the addresses defining $j$ occurs during the considered period. Since users can employ different wallets not necessarily linked by a transaction, the BUN should not be regarded as a perfect representation of the actual network of users but as an attempt at grouping addresses while minimising the presence of false positives.
 \vspace{-5mm}
\begin{table}[h!]
\centering
\caption{The four bubbles detected in Wheatley \textit{et al.} (2019).\cite{wheatley2019bitcoin}}
\begin{tabular}{c|c|c|c}
\hline
\hline
Bubble & Start & End & Days \\
\hline
\hline
1 & 25 May 2012 & 18 Aug 2012 & 84 \\
\hline
2 & 03 Jan 2013 & 11 Apr 2013 & 98 \\
\hline
3 & 07 Oct 2013 & 23 Nov 2013 & 47 \\
\hline
4 & 31 Mar 2017 & 18 Dec 2017 & 155\\
\hline
\hline
\end{tabular}
\vspace{5mm}
\label{tab:bubs}
\end{table}

\subsection{Data on Price Bubbles} As the valuation of cryptocurrencies is an emerging field of study, there is no consensus about the fundamental value of Bitcoin yet. As a consequence, understanding the meaning of a price bubble can be tricky. Here, we rely upon the methodology to identify bubbles developed by Wheatly \textit{et al.} (2019), employing a generalised version of Metcalfe's law to determine the fundamental value of Bitcoin and identifying four bubbles during the period under consideration (see Table \ref{tab:bubs}).\cite{wheatley2019bitcoin}

\section{Results}\label{sec:results}

\subsection{Analysis of Connected Components} Let us start by checking the connectedness of our weekly BUNs. In the earliest phase in the life of Bitcoin, \textit{i.e.}, until the fall of 2010, our BUNs are constituted by a large number of disconnected components, populated by few nodes (see Figure \ref{fig:cc_evolution}a). Such a large number can be explained by considering that the vast majority of nodes establish one-way transactions that bind vertices together only in a weak fashion. Besides, while the relative size of the Largest Weakly Connected Component (LWCC) has remained stable around $80\%$ of the total number of nodes throughout the entire period under consideration, the size of the Largest Strongly Connected Component (LSCC) has varied between $10\%$ and $30\%$ of the total number of nodes: specifically, a plateau of two years (2014 and 2015) during which the size of the LSCC has remained stable around $30\%$ of the total number of nodes can be observed (see Figure \ref{fig:cc_evolution}b). Lastly, the evolution of the ratio between the size of the LWCC/LSCC and the size of the second LWCC/LSCC reveals the former to be two-to-three orders of magnitude larger than the latter and comparable with the total size of the network (see Figure \ref{fig:cc_evolution}c)---a result further confirming that our BUNs are characterised by no more than one Largest Connected Component (LCC)---be it weakly or strongly connected.
\afterpage{%
\begin{figure*}[t!]
\centering
\subfloat[Number of disconnected components characterising our BUNs]{\includegraphics[width=\linewidth]{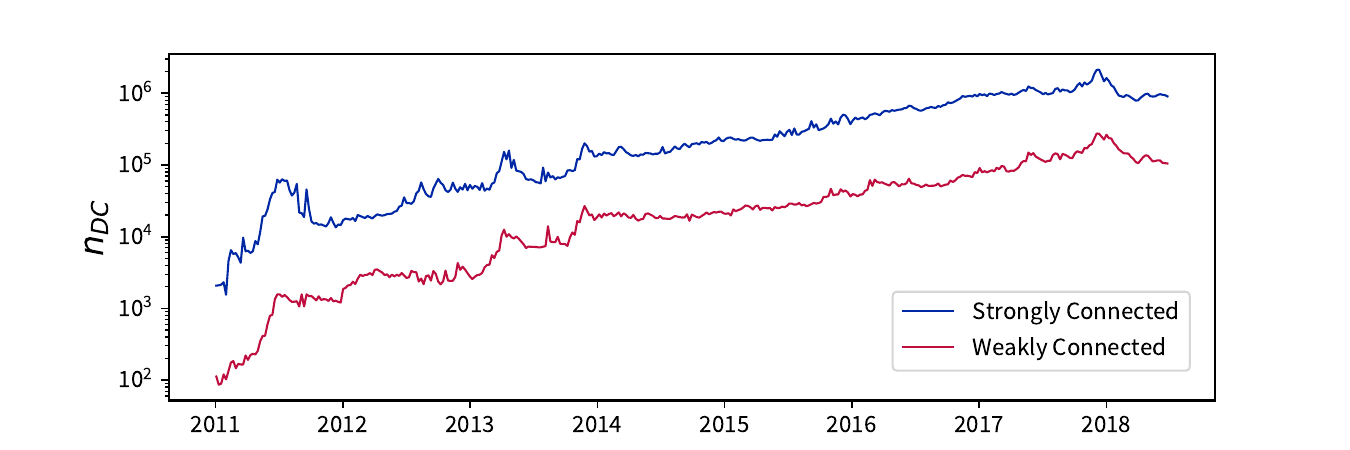}}

\subfloat[Relative size of the Largest Connected Component]{\includegraphics[width=\linewidth]{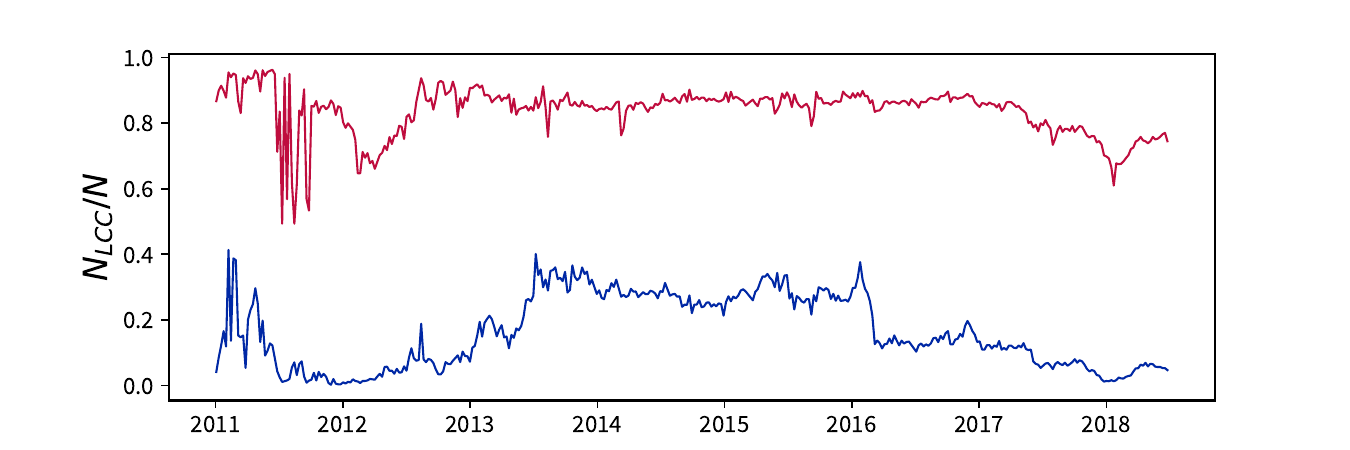}}

\subfloat[Ratio between the size of the LCC and of the size of the second LCC]{\includegraphics[width=\linewidth]{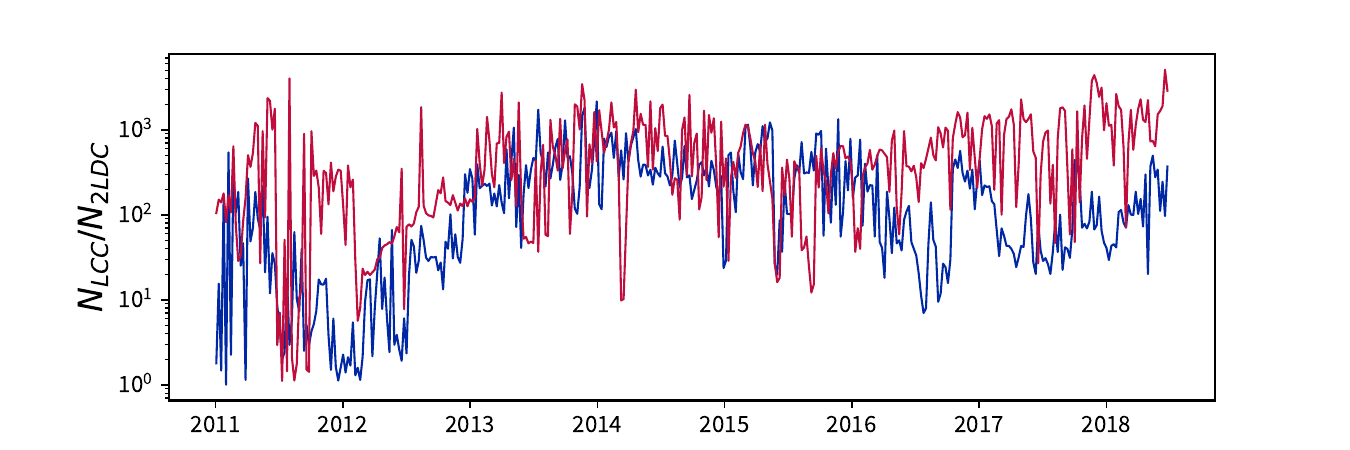}}
\caption{Panel (a) depicts the evolution of the number of weakly (red line) and strongly (blue line) disconnected components. Panel (b) depicts the evolution of the relative size of the LWCC (red line) and of the SWCC (blue line). Panel (c) depicts the evolution of the ratio between the size of the LCC and the size of the second LCC in the weak case (red line) and in the strong case (blue line). All measures are computed on the BUNs constructed on a weekly basis from 2011 to 2018.\vspace{5mm}}
\label{fig:cc_evolution}
\end{figure*}
\clearpage
}
\subsection{Analysis of Assortativity} 
\begin{figure*}[h!]
\centering
\includegraphics[width=\textwidth]{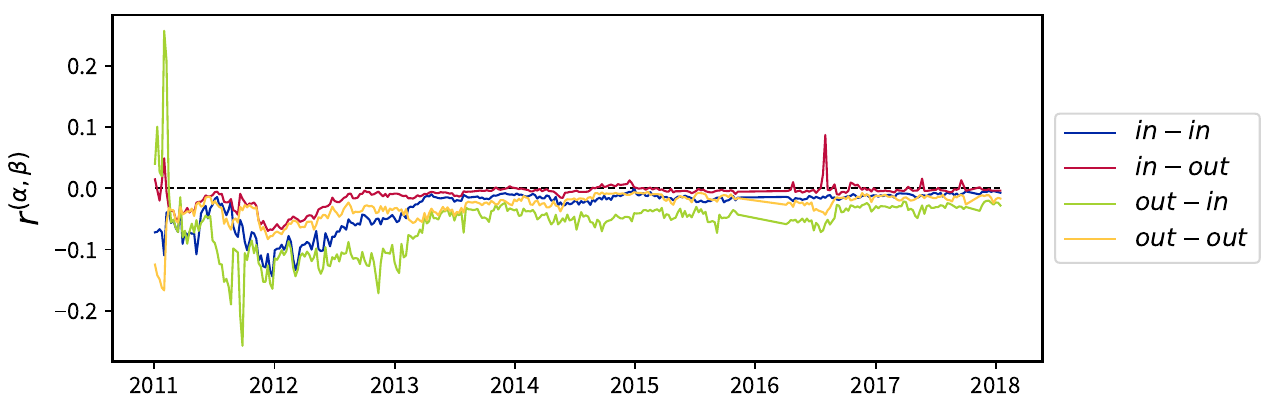}
\vspace{-5mm}
\caption{Evolution of the four directed variants of Newman's assortativity coefficient, revealing the weakly disassortative character of our BUNs.
}
\label{fig:degree_assortativity_r}
\end{figure*}

A network is said to be assortative when nodes with large (small) degrees tend to establish connections with nodes with large (small) degrees; disassortativity, instead, indicates the tendency of nodes with large (small) degrees to establish connections with nodes with small (large) degrees. Following Newman (2003) and Noldus and Van Mieghem (2015),\cite{newman2003mixing,noldus2015assortativity} the undirected assortativity coefficient is defined as

\begin{equation}\label{eq:newman_correlation}
r=\frac{\sum_{j,k}jk(e_{jk}-q_jq_k)}{\sigma^2_q}
\end{equation}
where the sum runs over the ``excess degree'' of a node---imagine entering a vertex following a specific edge: then, its ``excess degree'' is the degree of the vertex minus the edge we have followed. To be specific: $q_k \propto p_{k+1}$ is the ``excess degree'' probability distribution (with $p$ being the plain degree distribution), $\sigma^2_q$ is its standard deviation and $e_{jk}$ is the fraction of edges connecting nodes of degree $j$ with nodes of degree $k$. Naturally, $\sum_je_{jk}=q_k$. When considering directed networks, instead, four variants of the aforementioned coefficient can be calculated, \textit{i.e.}, the ones accounting for the correlation between out-degrees and out-degrees, out-degrees and in-degrees, in-degrees and out-degrees, and in-degrees and in-degrees. For example,

\begin{equation}
r^{(out,in)}=\frac{\sum_{j,k}jk(e_{jk}-q_j^{out}q_k^{in}) }{\sigma_{q^{out}}\sigma_{q^{in}}}
\end{equation}
where $e_{jk}$, now, represents the percentage of edges starting from nodes whose out-degree is $j$ and ending on nodes whose in-degree is $k$. Naturally, $\sum_je_{jk}=q_k^{in}$. In general:

\begin{equation}
r^{(\alpha,\beta)}=\frac{\sum_{j,k}jk(e_{jk}-q_j^{\alpha}q_k^{\beta}) }{\sigma_{q^{\alpha}}\sigma_{q^{\beta}}}
\end{equation}
where $e_{ij}$ is, now, the probability that a randomly chosen directed edge connects a vertex of $\alpha$-degree $j$ with a vertex of $\beta$-degree $k$, where $\alpha,\beta\in\{in,out\}$. Plotting the evolution of the aforementioned coefficients reveals the weakly disassortative character of our BUNs (see Figure \ref{fig:degree_assortativity_r}). More specifically, since $r^{out/in}$ asymptotically vanishes, one can conclude that $e_{jk}$ is increasingly close to $q_j^{out}q_k^{in}$, and it is analogously the same for the other indices.
\vspace{5mm}

\subsection{Bow-Tie Structure} Our BUNs are characterised by a mesoscopic kind of organisation known as \emph{bow-tie structure}. The definition of ``bow-tieness'' rests upon the concept of \emph{reachability}: we say that $j$ is \textit{reachable} from $i$ if a path from $i$ to $j$ exists. A directed graph is said to be \textit{strongly connected} if any two nodes are mutually reachable. Mutual reachability is an equivalence relation on the vertices of a graph, the equivalence classes being the strongly connected components of the graph itself. Hence, the bow-tie decomposition of a graph consists of the following sets of nodes:\cite{de2018reconstructing} 

\begin{itemize}
\item{} \textit{Strongly Connected Component (SCC)}: Each node within SCC can be reached by any other node within SCC;
\item{} \textit{In-Component}: Each node within IN can reach any other node within SCC, \textit{i.e.},
\begin{equation*}
\text{In}\equiv\{i\in V\setminus\text{SCC}\:|\:\text{SCC is reachable from $i$}\};
\end{equation*}
\item{} \textit{Out-Component}: Each node within OUT can be reached by any other node within SCC, \textit{i.e.},
\begin{equation*}
\text{Out}\equiv\{i\in V\setminus\text{SCC}\:|\:\text{$i$ is reachable from SCC}\};
\end{equation*}
\item{} \textit{Tubes}: Nodes not in the SCC from which you can reach both the In-Component and the Out-Component, \textit{i.e.},
\begin{equation*}
\text{Tubes} \equiv\{i\in V\setminus\text{SCC}\:\cup\:\text{IN}\cup\:\text{OUT}\:|\:\text{$i$ is reachable from IN and OUT is reachable from $i$}\}
\end{equation*}
\item{} \textit{In-Tendrils}: Sets of nodes that are reachable from the In-Component but do not reach the Out-Component, \textit{i.e.},
\begin{equation*}
\text{In-Tendrils} \equiv\{i\in V\setminus\text{SCC}\:|\:\text{$i$ is reachable from IN and OUT is not reachable from $i$}\};
\end{equation*}
\item{} \textit{Out-Tendrils}: Sets of nodes from which you can reach the Out-Component but that are not reachable from the In-Component, \textit{i.e.},
\begin{equation*}
\text{Out-Tendrils} \equiv\{i\in V\setminus\text{SCC}\:|\:\text{$i$ is not reachable from IN and OUT is reachable from $i$}\};
\end{equation*}
\item{} \textit{Others}: All nodes not belonging to any of the previous sets,
\begin{equation*}
\text{Other} \equiv\{i\in V\setminus\text{SCC}\:\cup\:\text{IN}\:\cup\:\text{OUT}\:\cup\:\text{Tubes}\:\cup\:\text{In-Tendrils}\:\cup\:\text{Out-Tendrils}\}.
\end{equation*}
\end{itemize}
\vspace{3mm}

\begin{figure*}[h!]
\centering
\includegraphics[width=\textwidth]{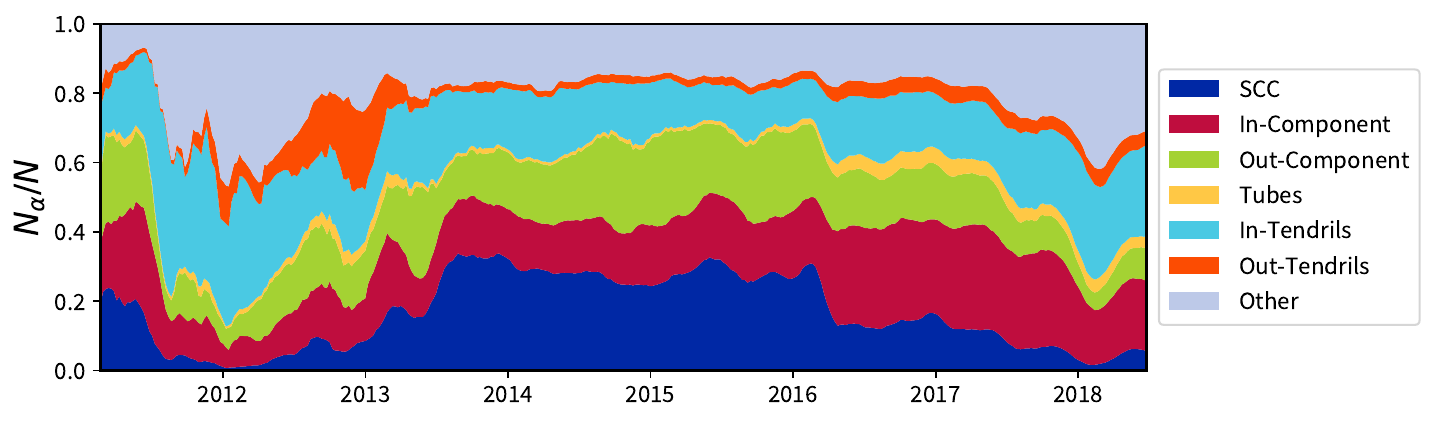}
\vspace{-5mm}
\caption{Evolution of the percentage of nodes belonging to each component of the bow-tie structure characterising our BUNs.
\vspace{-3mm}}
\label{fig:bow_tie}
\end{figure*}

As Figure \ref{fig:bow_tie} confirms, an SCC starts emerging in 2012, ``stabilizes'' around mid-2013 and persists until the end of 2015. Specifically, the SCC steadily rises during the biennium 2012-2013 and reaches $\simeq 30\%$ of the network size; afterwards, during the biennium 2014-2015, its size remains quite constant; then, during the last two years covered by our data set (\textit{i.e.}, 2016 and 2017), it shrinks and the percentage of nodes belonging to it goes back to the pre-2012 values. While during the biennium 2014-2015 the percentage of nodes constituting the SCC is larger than the percentage of nodes belonging to the other components, this is no longer true since 2016: in fact, while both the SCC and the Out-Component shrink, the In-Component becomes the dominant portion of the network.\cite{note1}\
\vspace{5mm}

\subsection{Core-Periphery Structure} 
\begin{figure}[h!]
\centering
\includegraphics[width=\textwidth]{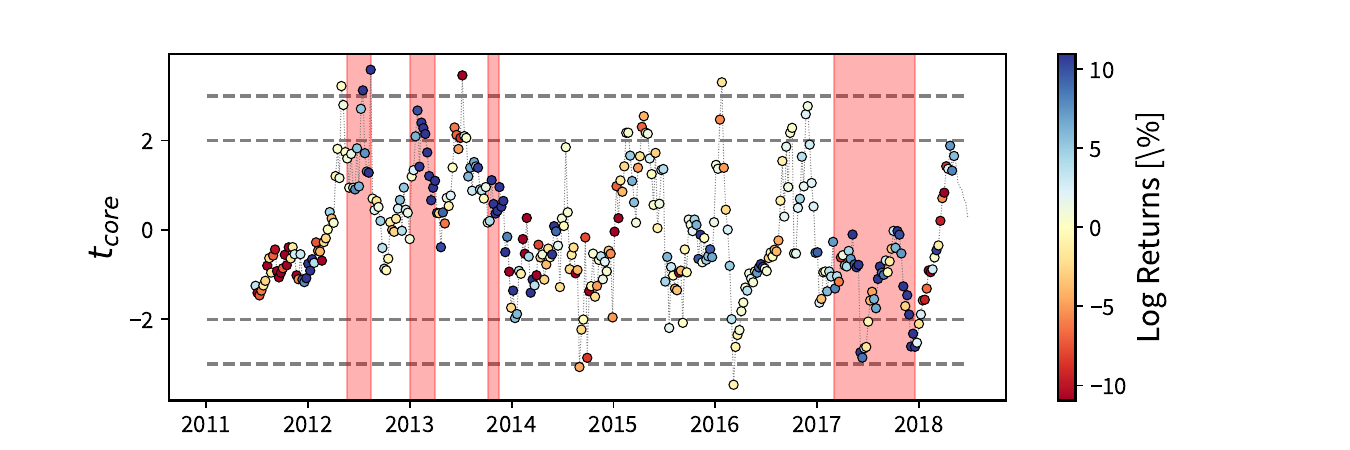}
\vspace{-5mm}
\caption{Evolution of the $t$-score of the percentage of nodes in the core, as defined by Equation \ref{eq:time-z-score}. The sample average and standard deviation have been computed over the set of empirical values covering six and half months before time $t$. Peaks are clearly visible in correspondence of the shaded areas identifying the bubbles, thus indicating the existence of periods during which the price and the structural quantities of interest co-evolve. Colours represent the log-returns of the Bitcoin price in USD, expressed as percentages and calculated over a rolling window of four weeks.\vspace{-5mm}}
\label{fig:zscore_core}
\end{figure}

Let us, now, inspect the degree of ``core-peripheryness'' characterising our BUNs. To this aim, we have run a recently-proposed method based on the multivariate extension of the \emph{surprise} score function (see also Appendix A).\cite{de2019detecting,marchese2022detecting} More precisely, applying the multivariate surprise to the partition induced by the bow-tie structure reveals the latter to induce a statistically significant core-periphery structure as well,\cite{note2} with the core coinciding with the SCC and the periphery gathering any other node. As expected from the results concerning the SCC, and confirmed by Figures \ref{fig:cc_evolution} and \ref{fig:zscore_core}, the periphery contains the vast majority of nodes throughout the first half of the Bitcoin history.

In order to gain insight into the correlations between the evolution of purely topological quantities and the Bitcoin price, let us plot the trend of the $t$-score defined as

\begin{equation}\label{eq:time-z-score}
t_X=\frac{X_t-\overline{X}}{s_X}
\end{equation}
for a generic quantity $X$ evaluated at time $t$, where the sample average $\overline{X}=\sum_t X_t/T$ and the sample standard deviation $s_X=$ \scalebox{0.6}{$\sqrt{\overline{X^2}-\overline{X}^2}$} have been computed over the set of values covering six and half months before time $t$.\cite{note3} As Figure \ref{fig:zscore_core} shows, the evolution of the $t$-score for the number of core and periphery nodes reveals the presence of peaks in correspondence withhe first three bubbles (identified by the shaded areas), thus indicating the existence of periods during which the price and the structural quantities of interest co-evolve. In particular, the number of nodes within the core and the periphery rises significantly in correspondence with the first three bubbles with respect to the temporal interval chosen as a benchmark, while the same quantities are characterised by a steep decrease during the periods between the bubbles.

\begin{figure}[b!]
\centering
\includegraphics[width=\linewidth]{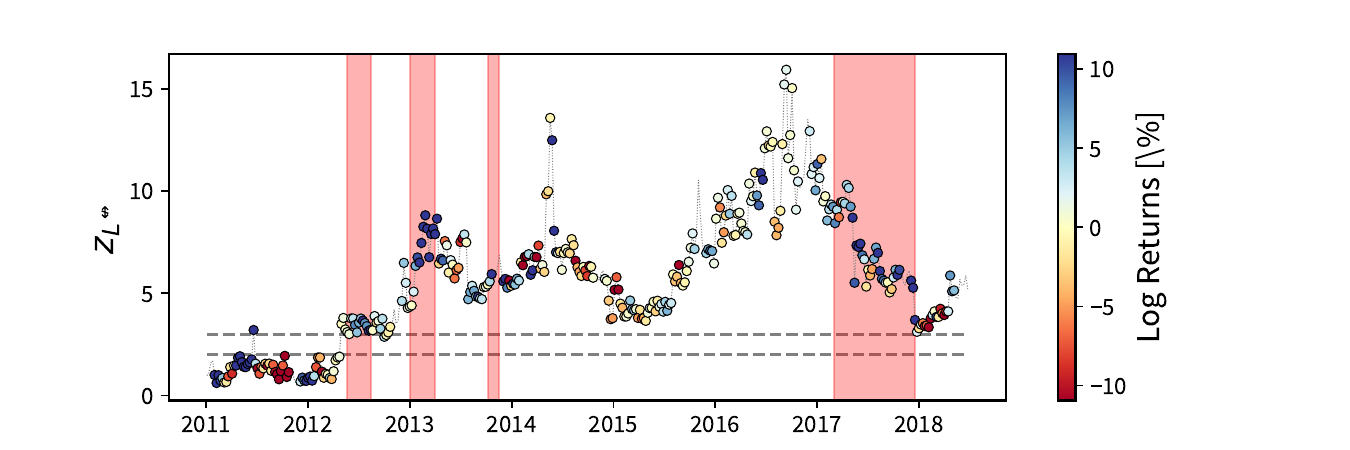}\\
\includegraphics[width=\linewidth]{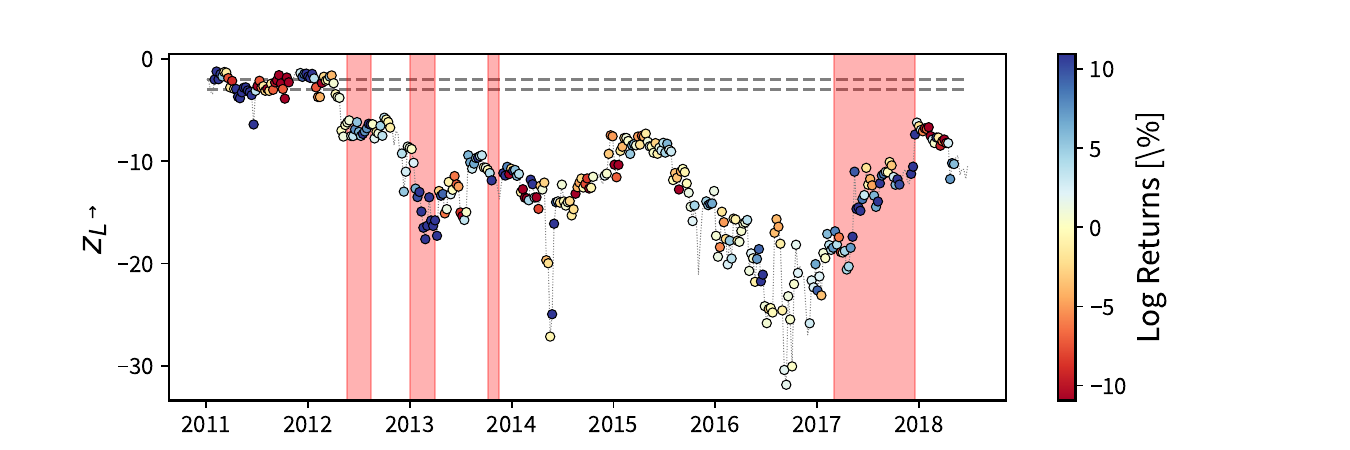}\\
\includegraphics[width=\linewidth]{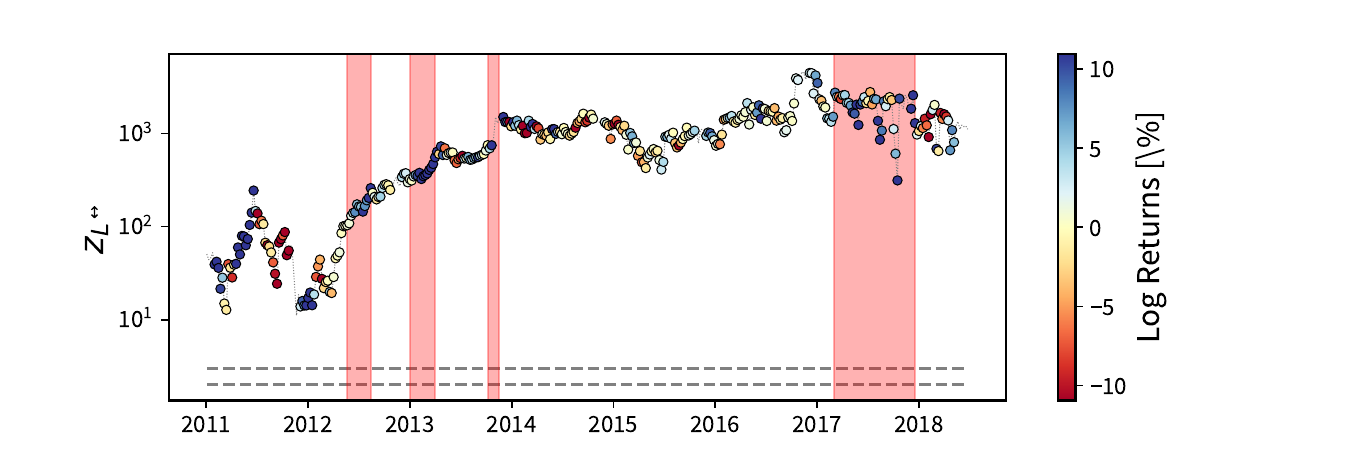}
\caption{Evolution of the $z$-score of the number of empty (top panel), non-reciprocated (middle panel) and reciprocated (bottom panel) dyads, computed over the ensemble induced by the DBCM. Dashed gray lines signal the values $z=\pm2$ and $z=\pm3$. The large positive values of our $z$-scores point out that observing links pointing in opposite directions is very unlikely---understandable by considering the level of sparseness of our BUNs; interestingly, events like these are more likely to occur in correspondence of the shaded areas identifying the bubbles, \textit{i.e.}, periods of price surge. Colours represent the log-returns of the Bitcoin price in USD, expressed as percentages and calculated over a rolling window of four weeks.\vspace{-5mm}}
\label{fig:dyads}
\end{figure}

\subsection{Dyadic Motifs} Let us now consider the quantities known as \emph{dyadic motifs} and the capturing of information concerning the patterns involving pairs of nodes.\cite{Squartini:2013,squartini2015stationarity} Three different patterns can be defined, \textit{i.e.},

\begin{itemize}
\item a \textit{reciprocated dyad}, indicating that both the connection $i\rightarrow j$ and the connection $j\rightarrow i$ exist. The total number of reciprocated dyads is quantified by calculating

\begin{equation}
L^{\leftrightarrow}=\sum_{i=1}^N\sum_{j(\neq i)} a_{ij}a_{ji};
\end{equation}
\item a \textit{non-reciprocated dyad}, indicating that either the connection $i\rightarrow j$ or the connection $j\rightarrow i$ exists. The total number of non-reciprocated dyads is quantified by calculating

\begin{equation}
L^{\rightarrow}=\sum_{i=1}^N\sum_{j(\neq i)} a_{ij}(1-a_{ji});
\end{equation}
\item an \textit{empty dyad}, indicating that neither the connection $i\rightarrow j$ nor the connection $j\rightarrow i$ exist. The total number of empty dyads is quantified by calculating

\begin{equation}
L^{\nleftrightarrow}=\sum_{i=1}^N\sum_{j(\neq i)} (1-a_{ij})(1-a_{ji}).
\end{equation}
\end{itemize}

We may study the emergence of dyadic motifs by adopting the following approach: instead of calculating the $t$-score to spot ``temporal outliers'', \textit{i.e.}, values that are statistically significant with respect to a series of precedent values, let us start by considering the $z$-score defined as
\begin{equation}\label{eq:zscore_sample}
z_X=\frac{X^*-\langle X\rangle}{\sigma_X}.
\end{equation}
We then quantify the extent to which the empirical value of the quantity $X^*\equiv X(\mathbf{A}^*)$, evaluated on the observed network $\mathbf{A}^*$, differs from its expectation, $\langle X \rangle$, evaluated by employing a specific benchmark; naturally, $\sigma_X$ represents the standard deviation of the quantity $X$, evaluated on the ensemble induced by the chosen benchmark. The $z$-score admits a clear, statistical interpretation: values satisfying $|z_X|>3$ signal that the empirical value is significantly over- or under-represented in the data, hence not compatible with the explanation solely provided by the constraints defining the chosen benchmark, at the $1\%$ level of statistical significance; on the other hand, values satisfying $|z_X|\leq3$ signal that the empirical value is compatible with the explanation solely provided by the constraints defining the chosen benchmark, at the $1\%$ level of statistical significance.

Hereby, we will employ the Exponential Random Graph Model named the Directed Binary Configuration Model (DBCM) to analyse the occurrence of the dyadic motifs.\cite{Squartini:2013,garlaschelli2008maximum} The results are shown in Figure \ref{fig:dyads}. The large, positive values of the $z$-scores for the empty and reciprocated dyads point out that our BUNs are much sparser and much more reciprocated than expected under a benchmark solely constraining the out- and in-degree sequences. Besides, we observe that reciprocated and non-reciprocated dyads have opposite trends as the number of reciprocated (non-reciprocated) dyads increases (decreases) during the growth of the (first three) bubbles: in other words, two previously single links become coupled, hence destroying two, non-reciprocated dyads and creating both an empty dyad and a reciprocated one (see also Appendix B).

\subsection{Centrality and Centralisation} Moving to the study of the mesoscale structure, we have computed three different centrality measures on the undirected version of our BUNs, \textit{i.e.}, the \textit{degree} centrality, the \textit{closeness} centrality and the \textit{betweenness} centrality.\cite{rodrigues2019network,newman2018networks} The degree centrality of node $i$ is proportional to its degree; the closeness centrality of node $i$ is proportional to the inverse of the sum of the distances separating $i$ from all the other nodes; the betweenness centrality of node $i$ is proportional to the fraction of shortest paths crossing $i$.
To summarise the information provided by any vector of centrality measures, hereby indicated with $\{c_i\}_{i=1}^N$, let us compute the two, aggregated quantities known as the \emph{Gini coefficient} and the \emph{centralisation index}.\cite{morgan1962anatomy, crucitti2006centrality,freeman1978centrality}. The Gini coefficient quantifies the unevenness of the distribution of a certain quantity such as the income.\cite{dixon1987bootstrapping} It is defined as
\begin{equation}\label{eq:gini_def}
G_c=\frac{\sum_{i=1}^N\sum_j|c_i-c_j|}{2N\sum_{i=1}^N{c_i}}
\end{equation}
and ranges between $0$ and $1$. While a Gini coefficient of $0$ indicates perfect evenness, a Gini coefficient of $1$ indicates perfect unevenness.\cite{note4} The trends in Figure \ref{fig:deg_centralisation} return the picture of an overall stationary system, suggesting the presence of a set of well-connected nodes, seemingly crossed by a large percentage of paths, that are also well inter-connected. The small values of the Gini coefficient induced by the closeness centrality can, in fact, be explained by the presence of vertices whose ``position'' ensures the vast majority of nodes to be reachable quite easily. The decrease of all Gini coefficients after 10 April 2013 is probably due to the Mt. Gox downward spiral, which started in the same year and ended with its bankruptcy in 2014.\cite{note5} Since then, $G_k$ has remained stable around the value of $0.5$---which is (still) compatible with a very centralised configuration. The Gini coefficients induced by the closeness and betweenness centralities are peaked in correspondence with the third bubble.
\afterpage{%
\begin{figure}[t!]
\centering
\includegraphics[width=\textwidth]{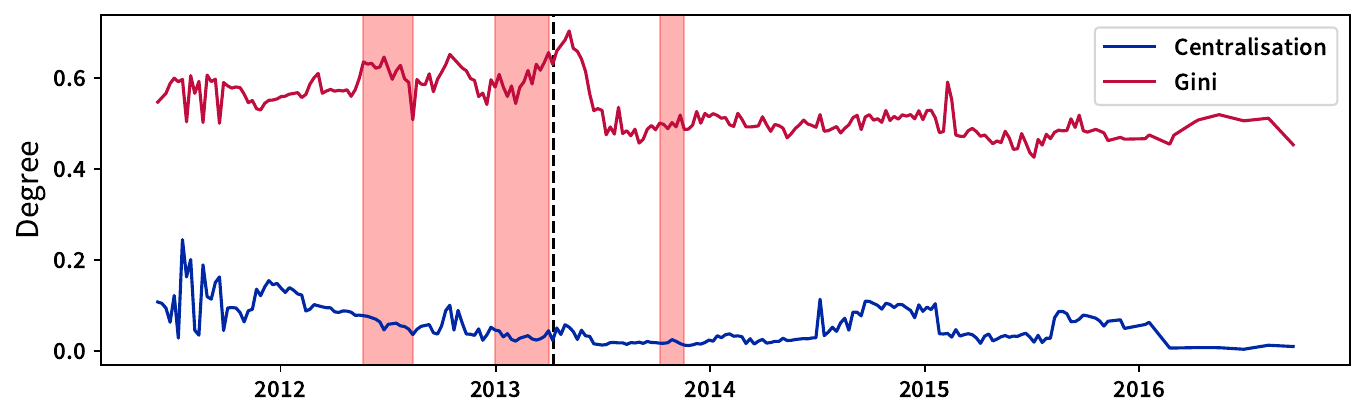}
\includegraphics[width=\textwidth]{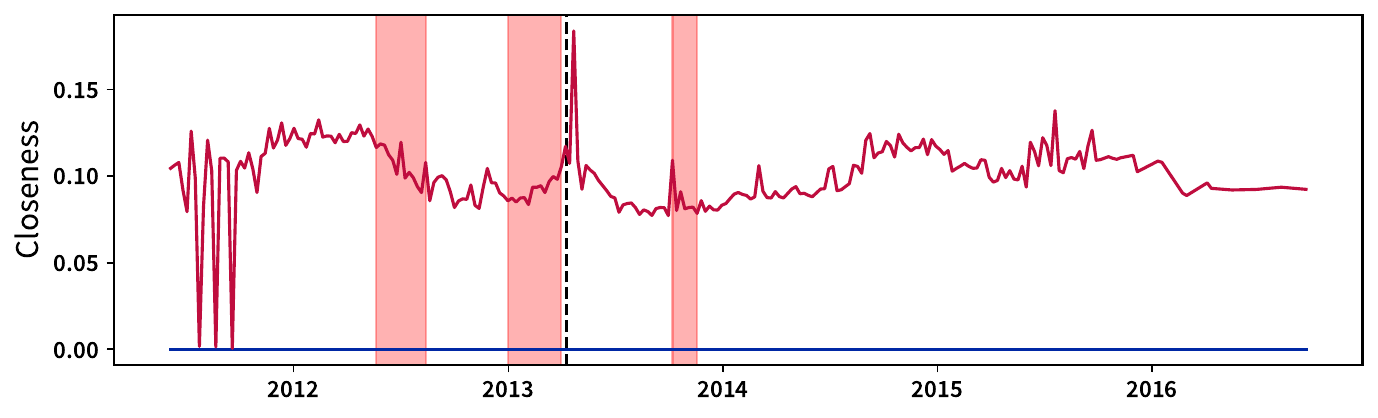}
\includegraphics[width=\textwidth]{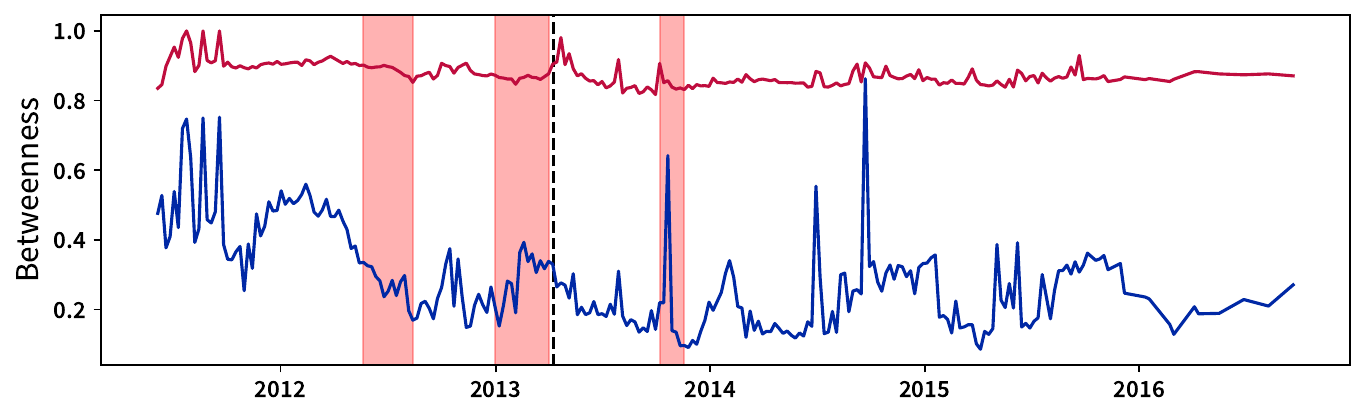}
\caption{Evolution of the Gini coefficient (in red) and centralisation index (in blue) for the degree, closeness and betweenness centrality measures. These trends return the picture of an overall stationary system, confirming the presence of a set of well-connected and interconnected nodes, crossed by a large percentage of paths. Shaded areas identify the bubbles. Co-movements between the price and the Gini coefficient of each centrality measure can be appreciated: particularly relevant is the spike observed on the 10th of April 2013, \textit{i.e.}, the day Mt. Gox suspended trading, thus triggering the burst of the price bubble, here indicated with a vertical, dashed line. Three spikes can be identified in 2011 as well: they are probably related to the small size of the network at the time and the instability caused by the Mt. Gox hacking in the summer of 2011.\vspace{5mm}}
\label{fig:deg_centralisation}
\end{figure}
\clearpage
}
Centralisation indices, instead, are intended to quantify the ``distance'' of a given configuration from the most unbalanced one. In mathematical terms, a generic centralisation index reads
\begin{equation}
C_c=\frac{\sum_{i=1}^N(c^*-c_i)}{\max\left\{\sum_{i=1}^N(c^*-c_i)\right\}}
\end{equation}
where $c^*=\max\{c_i\}_{i=1}^N$ represents the maximum value of the chosen centrality measured on the network under consideration and the denominator is calculated on the graph providing the maximum attainable value of the quantity $\sum_{i=1}^N(c^*-c_i)$ (see also Appendix C). For what concerns the degree, closeness and betweenness centrality, such a benchmark can be proven to be nothing else than a star graph with the same number of nodes of the network under inspection. More explicitly, in the case of the degree centrality, the centralisation index $C_k$ reads
\begin{equation}
C_k=\frac{\sum_{i=1}^N(k^*-k_i)}{(N-1)(N-2)}
\end{equation}
and a value $C_k=1$ would indicate that Bitcoin has become a star graph (at a certain point during its history). As Figure \ref{fig:deg_centralisation} reveals, the degree-induced centralisation has quickly stabilised around ``small'' values: contrarily to what the degree-induced Gini coefficient seems to suggest, Bitcoin is not evolving towards a star-like structure having a unique, central node participating to all transactions, but towards a structure where several hubs co-exist (see also our toy model in Appendix C). In other words, the (unrealistic) picture of a star-like structure can be replaced by the (more realistic) one depicting several locally star-like structures, their centers being vertices with a large number of connections, crossed by a large percentage of paths and well inter-connected (see also Figure \ref{fig:snaps}). As for the Gini coefficients, bubbles seem to have had some sort of impact on the evolution of the centralisation indices.
From a merely economic perspective, the presence of hubs increases the fragility of the Bitcoin ecosystem, making it more prone to large-scale losses and sudden price crashes in the event of failures.

\subsection{Small-Worldness} Finally, let us ask ourselves if Bitcoin satisfies the small-word property. To this aim, we have compared \emph{i)} the evolution of the \emph{average path length}
\begin{equation}
\text{APL}=\frac{\sum_{i=1}^N\sum_{j(>i)}d_{ij}}{N(N-1)/2}
\end{equation}
with the one predicted by the Erd\"os-R\'enyi Model (ERM), \textit{i.e.}, $\langle\text{APL}\rangle_\text{ERM}=\ln N/\ln\overline{k}$ where $\overline{k}=(N-1)p=2L/N$, and \emph{ii)} the evolution of the \emph{average clustering coefficient} (ACC):
\begin{equation}
\text{ACC}=\frac{\sum_{i=1}^NC_i}{N},
\end{equation}
where $C_i=\sum_{j(\neq i)}\sum_{k(\neq i,j)}a_{ij}a_{jk}a_{ki}/k_i(k_i-1)$, with the one predicted by the same model, \textit{i.e.}, $\langle\text{ACC}\rangle_\text{ERM}=p=2L/N(N-1)$. Computational constraints have forced us to limit our analysis to the giant connected component, during the first half of the period under consideration (\textit{i.e.}, until 2014). The results are displayed in Figure \ref{fig:small_world}: since $\ln(\ln N)<\text{APL}<\ln N\propto\langle\text{APL}\rangle_\text{ERM}$ and $\text{ACC}>\langle\text{ACC}\rangle_\text{ERM}$ across the entire period, Bitcoin can be claimed to be small-world.
\afterpage{%
\begin{figure}[h!]
\centering
\scalebox{0.8}{
\includegraphics[width=0.49\textwidth]{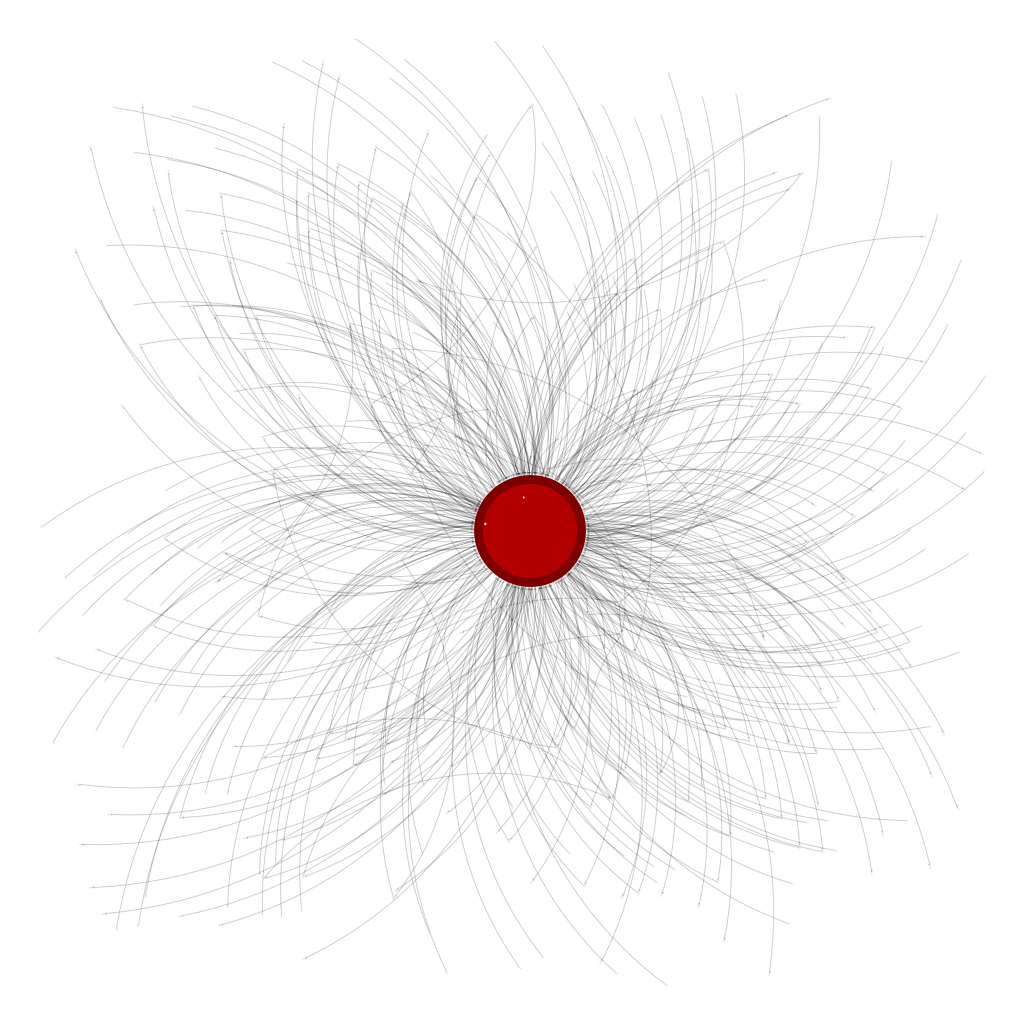}
\includegraphics[width=0.49\textwidth]{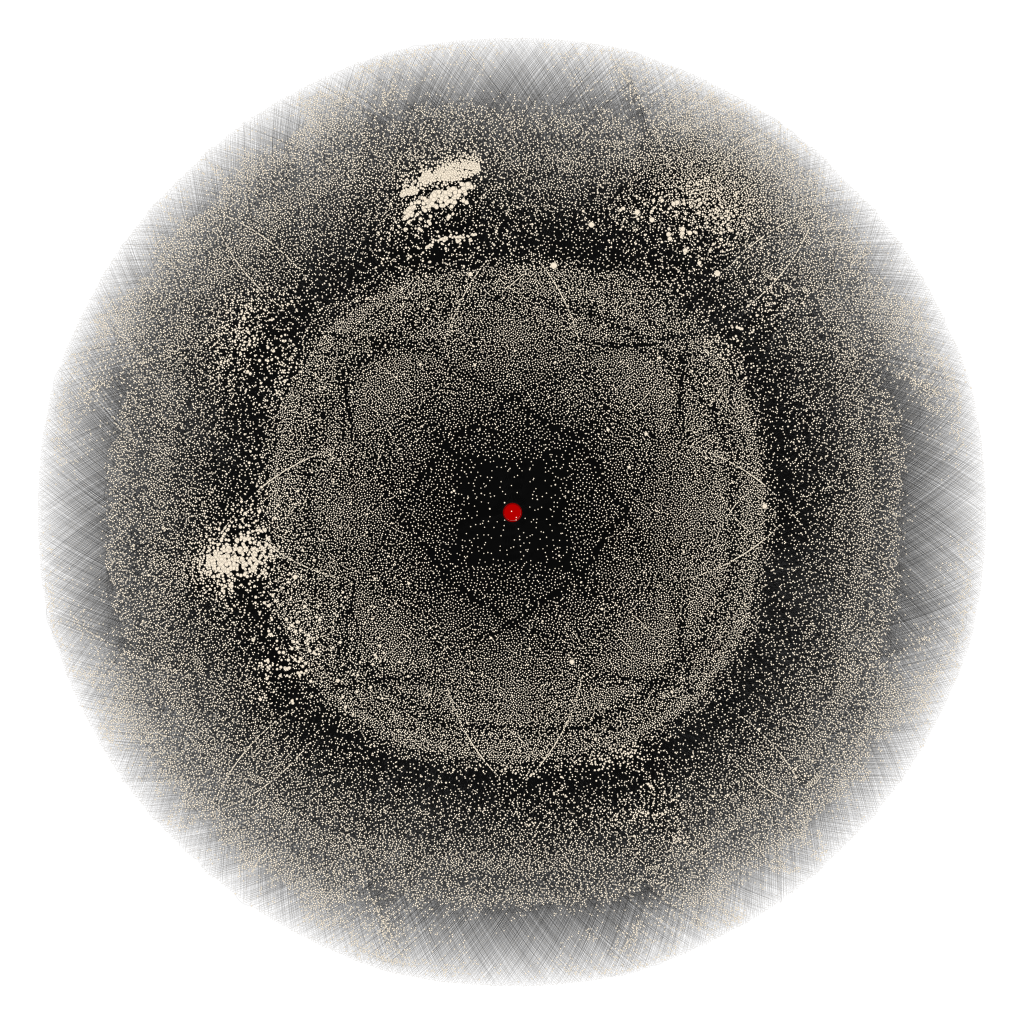}
}
\caption{Snapshots of Bitcoin BUN corresponding to 14 August 2010 (left) and 14 April 2013 (right). The two plots depict the ego network of the node having the largest degree. \vspace{2mm}}
\label{fig:snaps}
\end{figure}
\begin{figure}[h!]
\centering
\includegraphics[width=\textwidth]{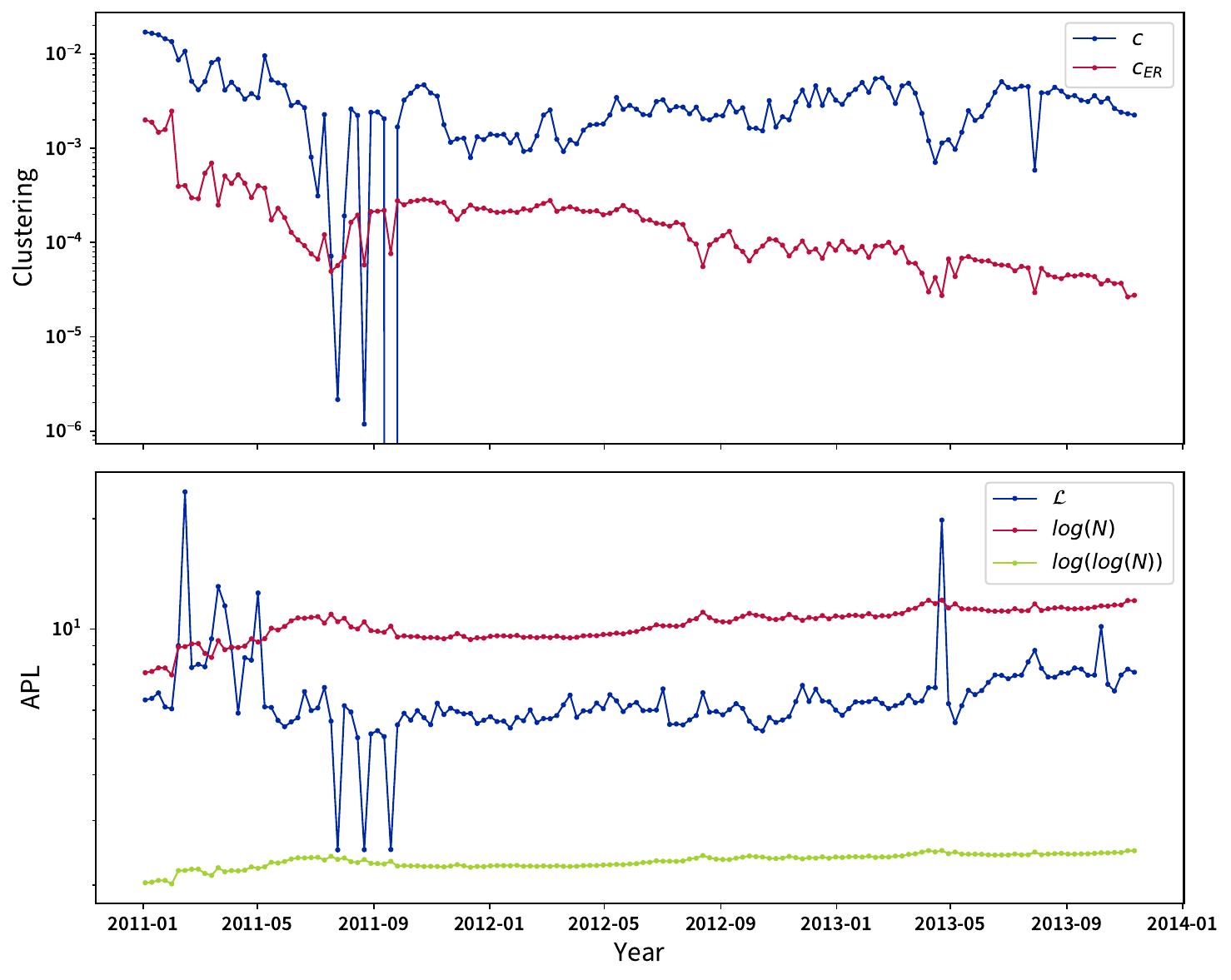}
\vspace{-8mm}
\caption{Evolution of the average path length and of the average clustering coefficient and their comparison with the corresponding values expected under the Erd\"os-R\'enyi Model (ERM): since $\ln(\ln N)<\text{APL}<\ln N\propto\langle\text{APL}\rangle_\text{ERM}$ and $\text{ACC}>\langle\text{ACC}\rangle_\text{ERM}$ across the entire period, Bitcoin can be claimed to be small-world.\vspace{-5mm}}
\label{fig:small_world}
\end{figure}
\clearpage
}

More in general, our BUNs tend to be poorly clustered, \textit{i.e.}, it is unlikely to observe two neighbours of a node that are also connected with each other. Still, the empirical value of the ACC is steadily one order of magnitude larger than the expected value of the ACC under the ERM, an evidence indicating that Bitcoin users tend to create many more triangles than those expected under a homogeneous benchmark induced by the link density.

\section{Conclusions}\label{sec:discussion}

In the present work, we have analysed the Bitcoin structural properties at the mesoscale and inspected their relationships with the price bubbles observed throughout the entire Bitcoin history. Features like the abundance of dyadic motifs and the values of the Gini coefficients have been observed to show significant variations in correspondence with (at least some of) the bubbles.\cite{note6} Some of the most representative examples are provided by the results concerning the analysis of centrality: during the third bubble, each trend is characterised by a peak probably due to the prominent role played by agents acting as liquidity providers, and likely also due to the exchange markets serving as structural hubs.

Although Bitcoin bubbles are difficult to forecast, our results suggest how to reduce the likelihood of them happening---or, at least, how to mitigate their impact: policies should be introduced to reduce the ``importance'' of hubs in the system. If only a few hubs account for most of the traffic in the network, in fact, the failure of any of them, at any point in time, has the potential to trigger the crash of a huge portion of the network itself. This is exactly what happened on 10 April 2013 when Mt. Gox, the major exchange market at the time, broke under the high trading volume.

\section*{Acknowledgements}

C.J.T. acknowledges financial support from the University of Zurich through the University Research Priority Program on Social Networks.

\section*{Author Contributions}

N.V. performed the analysis. N.V., T.S. and C.J.T. designed the research. All authors wrote, reviewed and approved the manuscript.

\section*{Conflict of Interest}
The authors declare that they have no known conflicts of interest as per the journal’s Conflict of Interest Policy.

\bibliographystyle{unsrt}  
\bibliography{main}  

\begin{thebibliography}{10}
\bibitem{note1}
Different results have been reported in Maesa \textit{et al.} (2019), where the
  authors observe the dominance of the SCC over the other components across the
  years 2009-2015. Such a result may be explained by the fact that \emph{i)}
  the heuristic clustering is applied only until 2015; \emph{ii)} only the
  multi-input one is employed; \emph{iii)} transactions are aggregated every 55
  days.\cite{maesa2019bow}.
\bibitem{note2}
The p-value associated with the multivariate surprise is steadily below the
  threshold of $5\%$.
\bibitem{note4}
A Gini coefficient of $0$ indicates that all actors have exactly the same
  income or the same number of interactions with the others. A Gini coefficient
  of $1$ indicates that the whole income of a population is concentrated in the
  hands of a single individual. In the case of nodes, a strict parallelism
  cannot be drawn as they need to be connected---for instance, the Gini
  coefficient of a star graph reads $G_{star}=2(N-1)(N-2)/4N(N-1)\to1/2$ when
  $N\to+\infty$.
\end{thebibliography}



\clearpage

\appendix

\setcounter{section}{0}

\section{Detecting Mesoscale Structures by Surprise}

Originally proposed to detect communities,\cite{aldecoa2013surprise,aldecoa2013exploring,nicolini2016modular} the \emph{surprise} reads

\begin{equation}\label{eq1}
\mathscr{S}=\sum_{l\geq l_\bullet^*}^{\min\{L, V_\bullet\}}\frac{\binom{V_\bullet}{l}\binom{V_\circ}{L-l}}{\binom{V}{L}}
\end{equation}
where $V$ is the volume of the network (coinciding with the total number of node pairs, \textit{i.e.}, $V=N(N-1)$ in the directed case), $V_\bullet$ is the total number of intra-cluster pairs (\textit{i.e.}, the number of node pairs within the individuated communities), and $L$ is the total number of links and $l^*$ is the observed number of intra-cluster links (\textit{i.e.}, the number of links within the individuated communities). In other words, $\mathscr{S}$ is the p-value of a hypergeometric distribution, aimed at testing the statistical significance of a given partition. Such a distribution, in fact, describes the probability of observing $l$ successes in $L$ draws (without replacement) from a finite population of size $V$ that contains exactly $V_\bullet$ objects with the desired feature (in our case, being an intra-cluster pair), each draw being interpreted either as a ``success'' or as a ``failure'': the smaller such a probability, the ``better'' the individuated partition---in other words, the more surprising its observation.

When carrying out a community detection exercise, links are understood as belonging to two different categories, \textit{i.e.}, the ones \emph{within} the clusters and the ones \emph{between} the clusters. The surprise-based formalism, however, can be extended to detect \emph{bimodular} structures, a terminology intended to describe either \emph{bipartite} or \emph{core-periphery} structures. In this case, three different ``species'' of links are needed (\textit{e.g.}, the core, the periphery and the core-periphery ones): for this reason, we need to consider the multinomial version of the surprise,\cite{de2019detecting} reading
\begin{equation}\label{eq2}
\mathscr{S}_\parallel=\sum_{i\geq l_\bullet^*}\sum_{j\geq l_\circ^*}\frac{\binom{V_\bullet}{i}\binom{V_\circ}{j}\binom{V-(V_\bullet+V_\circ)}{L-(i+j)}}{\binom{V}{L}}.
\end{equation}
From a technical point of view, $\mathscr{S}_\parallel$ is the p-value of a multivariate hypergeometric distribution, describing the probability of $i+j$ successes in $L$ draws (without replacement) from a finite population of size $V$ that contains exactly $V_\bullet$ objects with a first, specific feature and $V_\circ$ objects with a second, specific feature, each draw being either a ``success'' or a ``failure''. This is analogous to the univariate case, $i+j\in [l_\bullet^*+l_\circ^*,\min\{L,V_\bullet+V_\circ]$. According to the interpretation proposed in de Jeude \textit{et al.} (2019), revealing the core-periphery structure by minimising the surprise amounts at individuating the partition that is least likely to be explained by invoking the Erd\"os-R\'enyi Model (ERM) than by invoking the Stochastic Block Model (SBM).\cite{de2019detecting} This, in turn, amounts to individuating subgraphs with very different link densities, a piece of evidence that cannot be explained by a model defined by just one global parameter (such as the ERM).

\clearpage

\section{Evolution of Dyadic Motifs}

\begin{figure}[h!]
\centering
\includegraphics[width=\linewidth]{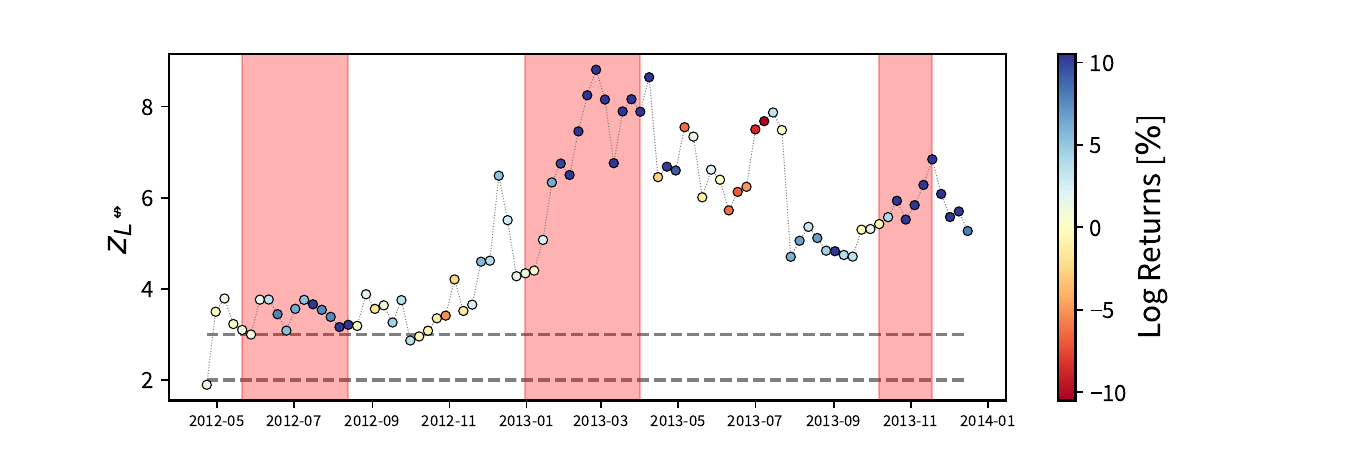}\\
\includegraphics[width=\linewidth]{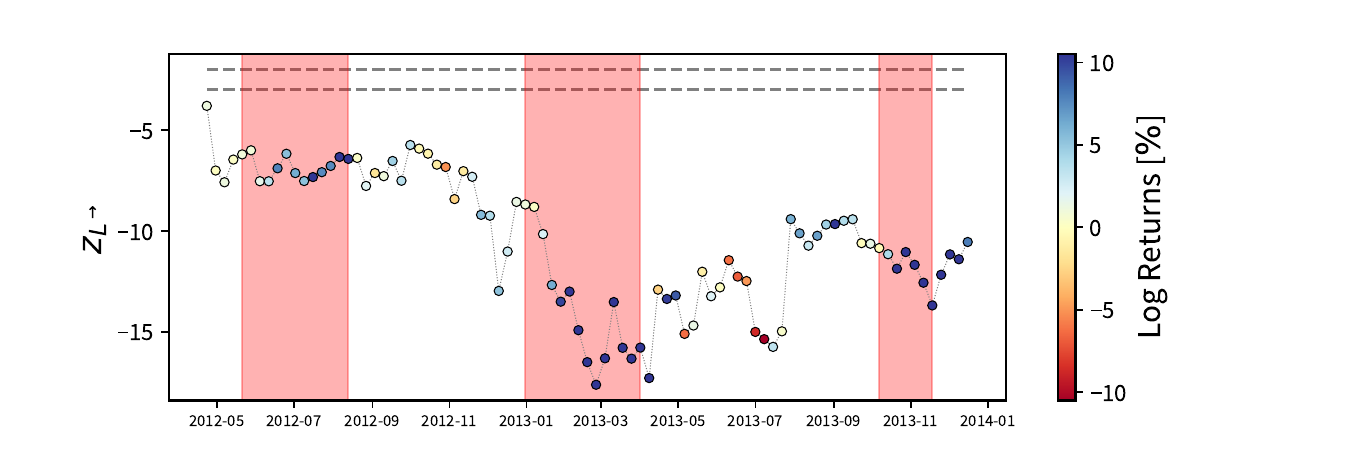}\\
\includegraphics[width=\linewidth]{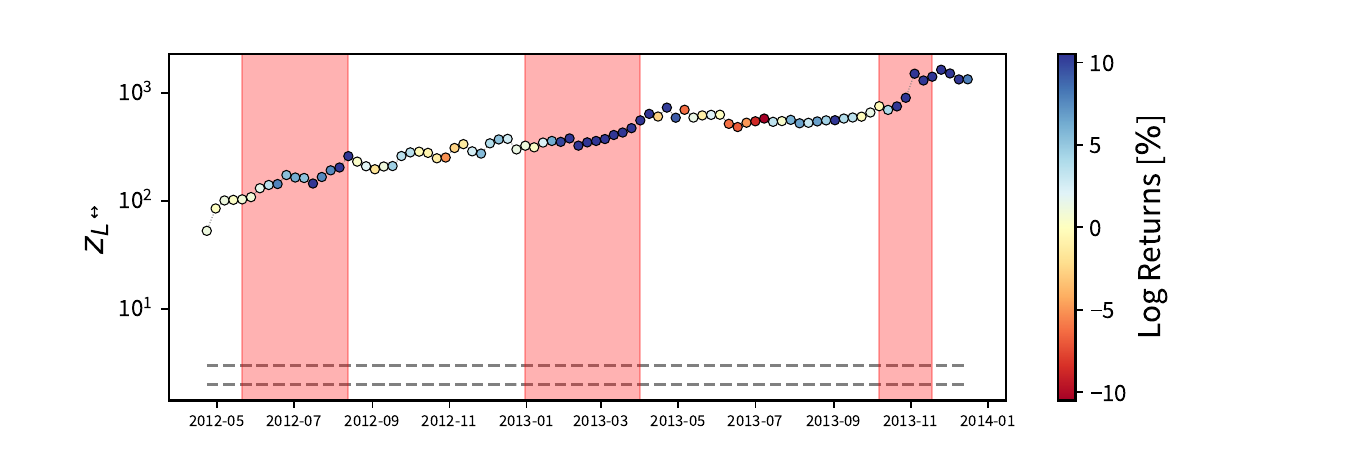}
\caption{Evolution of the $z$-score of the number of empty (top panel), non-reciprocated (middle panel) and reciprocated (bottom panel) dyads, computed over the ensemble induced by the DBCM, across the years 2012-2014. Dashed, gray lines signal the values $z=\pm2$ and $z=\pm3$. As already noticed in the main text, large, positive values for the $z$-score of the number of links pointing in opposite directions are more likely to occur in correspondence of the shaded areas identifying the bubbles, \textit{i.e.}, periods of price surge. Colours represent the log-returns of the Bitcoin price in USD, expressed as percentages and calculated over a rolling window of four weeks.\vspace{-5mm}}
\label{fig:dyads2}
\end{figure}

\clearpage

\section{Centralisation Indices}

\textit{C.1. Non-Normalised Centrality Measures}---The degree-induced centralisation index reads
\begin{equation}
C_k=\frac{\sum_{i=1}^N(k^*-k_i)}{(N-1)(N-2)}
\end{equation}
where $k_i$ is the non-normalised degree centrality of node $i$. Notice that a star graph with $N$ nodes has $N-1$ leaves linked to the central hub: hence, $k_h=k^*=N-1$, $k_l=1$ and the numerator reduces to $N-1$ addenda each one reading $N-1-1=N-2$.\cite{freeman1978centrality,lin2020lightning,lin2022weighted}\\

The closeness-induced centralisation index reads
\begin{equation}
C_c=\frac{\sum_{i=1}^N(c^*-c_i)}{(N-2)/(2N-3)}
\end{equation}
where $c_i$ is the non-normalised closeness centrality of node $i$. Notice that a star graph with $N$ nodes has $N-1$ leaves linked to the central hub: hence, $c_h=c^*=1/(N-1)$, $c_l=1/(1+2(N-2))=1/(2N-3)$ and the numerator reduces to $N-1$ addenda each one reading $(N-2)/(N-1)(2N-3)$.\cite{freeman1978centrality,lin2020lightning,lin2022weighted}\\

The betweenness-induced centralisation index reads
\begin{equation}
C_b=\frac{\sum_{i=1}^N(b^*-b_i)}{(N-1)^2(N-2)/2}
\end{equation}
where $b_i$ is the non-normalised betweenness centrality of node $i$. Notice that a star graph with $N$ nodes has $N-1$ leaves linked to the central hub: hence, $b_h=b^*=(N-1)(N-2)/2$, $b_l=0$ and the numerator reduces to $N-1$ addenda each one reading $(N-1)(N-2)/2$.\cite{freeman1978centrality,lin2020lightning,lin2022weighted}
\vspace{5mm}
\\
\textit{C.2. Normalised Centrality Measures}---The degree-induced centralisation index reads
\begin{equation}
C_k=\frac{\sum_{i=1}^N(\underline{k}^*-\underline{k}_i)}{N-2}
\end{equation}
where $\underline{k}_i$ is the normalised degree centrality of node $i$. Notice that a star graph with $N$ nodes has $N-1$ leaves linked to the central hub: hence, $\underline{k}_h=\underline{k}^*=1$, $\underline{k}_l=1/(N-1)$ and the numerator reduces to $N-1$ addenda each one reading $1-1/(N-1)=(N-2)/(N-1)$.\cite{freeman1978centrality,lin2020lightning,lin2022weighted}\\

The closeness-induced centralisation index reads
\begin{equation}
C_c=\frac{\sum_{i=1}^N(\underline{c}^*-\underline{c}_i)}{(N-1)(N-2)/(2N-3)}
\end{equation}
where $\underline{c}_i$ is the normalised closeness centrality of node $i$. Notice that a star graph with $N$ nodes has $N-1$ leaves linked to the central hub: hence, $\underline{c}_h=\underline{c}^*=1$, $\underline{c}_l=(N-1)/(1+2(N-2))=(N-1)/(2N-3)$ and the numerator reduces to $N-1$ addenda each one reading $(N-2)/(2N-3)$.\cite{freeman1978centrality,lin2020lightning,lin2022weighted}\\

The betweenness-induced centralisation index reads
\begin{equation}
C_b=\frac{\sum_{i=1}^N(\underline{b}^*-\underline{b}_i)}{N-1}
\end{equation}
where $\underline{b}_i$ is the normalised betweenness centrality of node $i$. Notice that a star graph with $N$ nodes has $N-1$ leaves linked to the central hub: hence, $\underline{b}_h=\underline{b}^*=1$, $\underline{b}_l=0$ and the numerator reduces to $N-1$ addenda each one reading $1$.\cite{freeman1978centrality,lin2020lightning}
\vspace{5mm}
\\
\textit{C.3. The Bitcoin Topology: A Toy Model}---A toy model can help reconciling the two, apparently contradictory, results provided by the Gini coefficient and the degree-induced centralisation. Imagine $N_h$ hubs connected between them and $N_l$ leaves connected to each of them; hence, the total number of nodes reads $N=N_h(N_l+1)$, the degree of each hub reads $k_h=(N_h-1)+N_l$, the degree of each leave reads $k_l=1$ and
\begin{align}
G_k=\frac{\sum_{i=1}^N\sum_j^N|k_i-k_j|}{2N\sum_{i=1}^N{k_i}}&=\frac{2N_h^2N_l[(N_h-1)+N_l-1]}{2N_h(N_l+1)\left[N_h(N_h-1)+2N_hN_l\right]}\nonumber\\
&=\frac{N_h^2N_l[(N_h-1)+N_l-1]}{N_h(N_l+1)\left[N_h(N_h-1)+2N_hN_l\right]}\nonumber\\
&=\frac{N_l(N_h+N_l-2)}{(N_l+1)\left[(N_h-1)+2N_l\right]}\simeq\frac{N_h+N_l}{N_h+2N_l};
\end{align}
now, $G_k\simeq2/3$ if $N_l=N_h$, \textit{i.e.}, if each hub is linked to a number of leaves that matches the total number of hubs and $G_k\rightarrow1/2$ as $N_l\rightarrow+\infty$ (\textit{i.e.}, if the number of leaves per hub becomes ``very large''). In this setting, we have that
\begin{equation}
C_k=\frac{\sum_{i=1}^N(k^*-k_i)}{(N-1)(N-2)}=\frac{N_hN_l[(N_h-1)+N_l-1]}{[N_h(N_l+1)-1][N_h(N_l+1)-2]}\simeq\frac{N_h+N_l}{N_hN_l}
\end{equation}
which amounts at $C_k\simeq 0.02$ if we set $N_h=N_l=100$. Hence, we can recover core-periphery structures for which a large Gini coefficient co-exists with a small degree-centralisation by opportunely tuning the parameters of our toy model.

\end{document}